\documentclass[twocolumn]{aastex63}

\newcommand{\MS}{\ifmmode{\,}\else\thinspace\fi{\rm M}\ifmmode_{\odot}\else$_{\odot}$\fi}
\newcommand{\LS}{\ifmmode{\,}\else\thinspace\fi{\rm L}\ifmmode_{\odot}\else$_{\odot}$\fi}
\newcommand{\RS}{\ifmmode{\,}\else\thinspace\fi{\rm R}\ifmmode_{\odot}\else$_{\odot}$\fi}
\newcommand{\teff}{\ifmmode T_{\rm eff}\else$T_{\rm eff}$\fi}
\newcommand{\Ke}{\ifmmode{\,}\else\thinspace\fi{\rm K}}
\received{February 24, 2021}
\revised{March 23, 2021}
\accepted{March 31, 2021}

\submitjournal{ApJL}

\shorttitle{Binarity as the Origin of Long Secondary Periods in Red Giant Stars}
\shortauthors{Soszy\'nski et al.}

\begin{document}

\title{Binarity as the Origin of Long Secondary Periods in Red Giant Stars}

\author{I. Soszy\'nski}
\affiliation{Astronomical Observatory, University of Warsaw, Al.~Ujazdowskie~4, 00-478~Warszawa, Poland}
\email{soszynsk@astrouw.edu.pl}

\author{A. Olechowska}
\affiliation{Astronomical Observatory, University of Warsaw, Al.~Ujazdowskie~4, 00-478~Warszawa, Poland}

\author{M. Ratajczak}
\affiliation{Astronomical Observatory, University of Warsaw, Al.~Ujazdowskie~4, 00-478~Warszawa, Poland}

\author{P. Iwanek}
\affiliation{Astronomical Observatory, University of Warsaw, Al.~Ujazdowskie~4, 00-478~Warszawa, Poland}

\author{D. M. Skowron}
\affiliation{Astronomical Observatory, University of Warsaw, Al.~Ujazdowskie~4, 00-478~Warszawa, Poland}

\author{P. Mr\'oz}
\affiliation{Astronomical Observatory, University of Warsaw, Al.~Ujazdowskie~4, 00-478~Warszawa, Poland}
\affiliation{Division of Physics, Mathematics, and Astronomy, California Institute of Technology, Pasadena, CA 91125, USA}

\author{P. Pietrukowicz}
\affiliation{Astronomical Observatory, University of Warsaw, Al.~Ujazdowskie~4, 00-478~Warszawa, Poland}

\author{A. Udalski}
\affiliation{Astronomical Observatory, University of Warsaw, Al.~Ujazdowskie~4, 00-478~Warszawa, Poland}

\author{M. K. Szyma\'nski}
\affiliation{Astronomical Observatory, University of Warsaw, Al.~Ujazdowskie~4, 00-478~Warszawa, Poland}

\author{J. Skowron}
\affiliation{Astronomical Observatory, University of Warsaw, Al.~Ujazdowskie~4, 00-478~Warszawa, Poland}

\author{M. Gromadzki}
\affiliation{Astronomical Observatory, University of Warsaw, Al.~Ujazdowskie~4, 00-478~Warszawa, Poland}

\author{R. Poleski}
\affiliation{Astronomical Observatory, University of Warsaw, Al.~Ujazdowskie~4, 00-478~Warszawa, Poland}

\author{S. Koz\l{}owski}
\affiliation{Astronomical Observatory, University of Warsaw, Al.~Ujazdowskie~4, 00-478~Warszawa, Poland}

\author{M. Wrona}
\affiliation{Astronomical Observatory, University of Warsaw, Al.~Ujazdowskie~4, 00-478~Warszawa, Poland}

\author{K. Ulaczyk}
\affiliation{Astronomical Observatory, University of Warsaw, Al.~Ujazdowskie~4, 00-478~Warszawa, Poland}
\affiliation{Department of Physics, University of Warwick, Gibbet Hill Road, Coventry, CV4~7AL,~UK}

\author{K. Rybicki}
\affiliation{Astronomical Observatory, University of Warsaw, Al.~Ujazdowskie~4, 00-478~Warszawa, Poland}

\begin{abstract}

Long secondary periods (LSPs), observed in a third of pulsating red giant stars, are the only unexplained type of large-amplitude stellar variability known at this time. Here we show that this phenomenon is a manifestation of a substellar or stellar companion orbiting the red giant star. Our investigation is based on a sample of about 16,000 well-defined LSP variables detected in the long-term OGLE photometric database of the Milky Way and Magellanic Clouds, combined with the mid-infrared data extracted from the NEOWISE-R archive. From this collection, we selected about 700 objects with stable, large-amplitude, well-sampled infrared light curves and found that about half of them exhibit secondary eclipses, thus presenting an important piece of evidence that the physical mechanism responsible for LSPs is binarity. Namely, the LSP light changes are due to the presence of a dusty cloud orbiting the red giant together with the companion and obscuring the star once per orbit. The secondary eclipses, visible only in the infrared wavelength, occur when the cloud is hidden behind the giant. In this scenario, the low-mass companion is a former planet that has accreted a significant amount of mass from the envelope of its host star and grown into a brown dwarf.

\end{abstract}


\section{Introduction}

As a red giant star evolves along the asymptotic giant branch (AGB) and it exceeds a certain luminosity, it becomes a variable star. The photometric variability of red giants manifests itself in several ways. Single or (more frequently) multiperiodic brightness changes are caused by stellar pulsations, irregular variations signal the presence of circumstellar dust, and stochastic noise in the light curves may result from the supergranular convection in the envelopes of giant stars. However, there is one type of red giant variability that has not yet found a satisfactory explanation -- the so-called long secondary periods (LSPs).

\begin{figure}
\centering
\includegraphics[width=8.3cm]{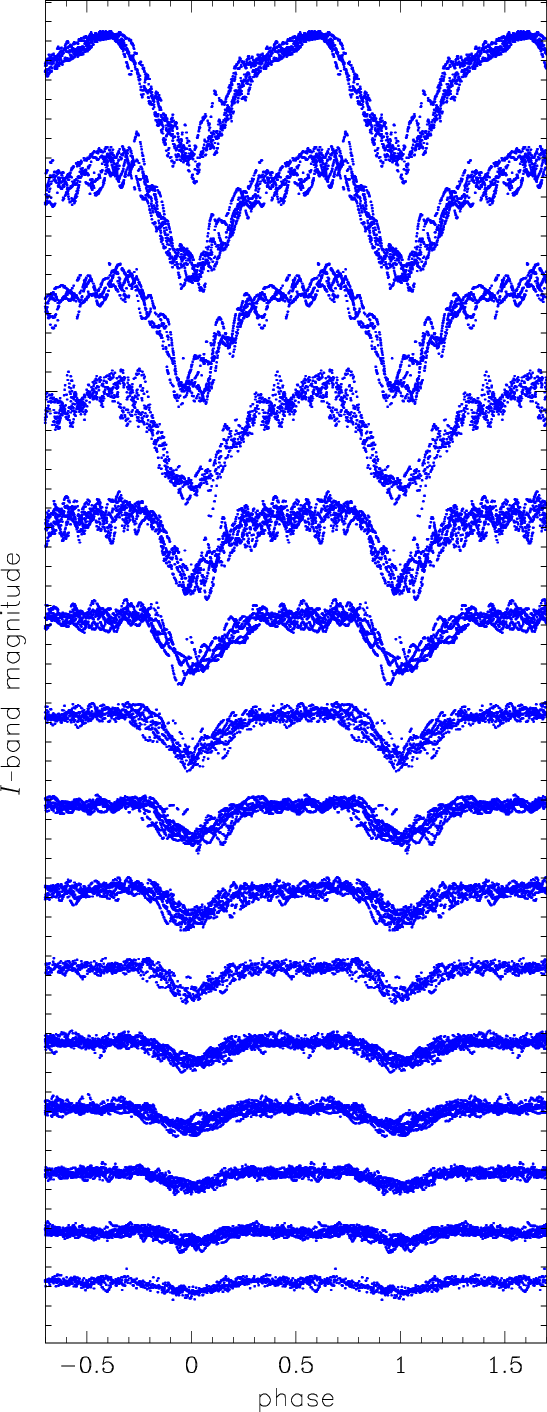}
\caption{Example OGLE {\it I}-band light curves of LSP variables from the Galactic bulge and LMC. Tick marks on the vertical axis are spaced at 0.1~mag intervals.\label{fig1}}
\end{figure}

LSPs range from several months to several years -- an order of magnitude longer than the pulsation periods. This phenomenon can be detected in at least a third of luminous AGB stars and supergiants. It is likely that low-amplitude LSP variations occur also in the brightest first-ascent red giant branch (RGB) stars\footnote{In this paper we use the term ``red giant'' for both AGB and RGB stars} \citep{pawlak2021}. The visual amplitudes of the LSP variations reach 1~mag, while the peak-to-peak radial velocity amplitudes range from 2 to 7~km~s$^{-1}$, with a clustering around 3.5~km~s$^{-1}$ \citep{hinkle2002,wood2004,nicholls2009}. In the period--luminosity plane, LSP variables obey a relation called sequence~D \citep{wood1999}, which partly overlaps with sequence~E populated by red giants in close binary systems showing ellipsoidal variability \citep{soszynski2004,nicholls2010}.

It is now beyond doubt that the presence of the LSPs in red giants is associated with unusual amounts of circumstellar dust. \citet{wood2009} found a mid-infrared (mid-IR) excess in the LSP variables compared to stars without LSPs, which indicates the presence of circumstellar matter absorbing the energy in the optical light and reemitting it at longer wavelengths. Moreover, \citet{wood2009} noticed that dust around LSP variables is not spherically symmetrically distributed, but it is rather in a clumpy or a disk-like configuration. \citet{mcdonald2019} linked the onset of the LSP phenomenon in AGB stars with the onset of strong mass-loss due to dust-driven winds. \citet{pawlak2021} analyzed spectral energy distributions of a large number of LSP and non-LSP variables and confirmed these conclusions.

Numerous authors have explored various scenarios for the origin of LSPs \citep[e.g.][]{hinkle2002,olivier2003,wood2004,derekas2006,nicholls2009,nie2010,stothers2010,takayama2015,percy2016}, but were unable to give a final solution to this problem. In the recent years, two the most favored explanations have been discussed in the literature: oscillatory convective nonradial modes \citep[e.g.][]{saio2015,takayama2020} and eclipses by a dusty cloud surrounding and following a close-orbiting brown-dwarf or low-mass stellar companion. The latter hypothesis was proposed by \citet{wood1999} and developed by \citet{soszynski2007} and \citet{soszynski2014}. In this scenario, the presence of the brown-dwarf companion is necessary to produce the observed radial velocity amplitudes, but the large fraction of the LSP variables among red giant stars can only be explained by the fact that the companion was initially an exoplanet that accreted a significant amount of mass from the envelope of its host. 

There have been surprisingly few papers that addressed the characteristic light-curve shapes of LSP variables. The LSP light curves have significantly different morphology than the light curves produced by both radial and nonradial pulsation modes in red giants. The sample of {\it I}-band light curves of 15 LSP variables from the Galactic bulge and Large Magellanic Cloud (LMC) are displayed in Figure~\ref{fig1}. The light curves are folded with the LSPs and sorted by their amplitudes. Here, we selected objects with relatively small amplitudes of the red giant pulsations (visible as short-period oscillations in the light curves) to highlight the characteristic features of the LSP modulation.

A typical LSP cycle can be divided into two parts, each lasting approximately half of the period. During the first part, the brightness does not vary at all (except pulsations) or it increases slowly with time. The second part of the light curve has a triangular shape -- the luminosity decreases, reaches a minimum, and then increases. In the largest-amplitude LSP variables, the duration of this triangle minimum may extend and exceed half the cycle, while the maxima often take a rounded shape. In some stars, the minima of the light curves significantly vary in depth from cycle to cycle, sometimes the LSP modulation completely disappears for a period of time.

\citet{soszynski2014} demonstrated that the LSP light curves can be produced by a dusty cloud with a comet-like tail trailing behind a low-mass companion. It is worth noting that although the radial velocity amplitudes in the vast majority of LSP variables are consistent with the presence of a brown-dwarf companion, the characteristic LSP light curves can also be observed in stellar binary systems, for example, in some symbiotic stars \citep[e.g.][]{gromadzki2013}. Additionally, \citet{soszynski2014} showed that the nonsinusoidal shapes of the radial velocity curves \citep[e.g.][]{hinkle2009,nicholls2009} can be explained by the Rossiter-McLaughlin effect, even if a companion is in a circular orbit. On the other hand, the stellar pulsation models are capable of obtaining periods \citep[e.g.][]{saio2015} or amplitudes \citep[e.g.][]{takayama2020} of the LSP modulation, but are unable to reproduce the characteristic light-curve shapes shown in Figure~\ref{fig1}.

In this paper, we study the mid-infrared (mid-IR) light curves of LSP giants detected in the Optical Gravitational Lensing Experiment (OGLE) photometric databases. The binary model of the LSP phenomenon predicts an additional effect that should be visible in the mid-IR regime -- the secondary eclipses that occur when the dusty cloud hides behind the red giant. The discovery of the secondary eclipses in the LSP light curves would be a strong argument supporting the binary explanation of the LSPs in red giant stars.

\section{Sample Selection}

The OGLE project \citep{udalski2015} is an ongoing sky survey conducting photometric observations of the Magellanic Clouds and the Galactic bulge and disk, about 3500 square degrees in total. The observations are carried out with the 1.3~m Warsaw Telescope at Las Campanas Observatory, Chile. The OGLE database contains long-term light curves for about 2~billion stars obtained in the {\it I-} and {\it V}-band passbands from the Johnson--Cousins photometric system. This dataset is ideally suited to search for variable red giants and study their photometric properties.

Based on observations obtained between 1997 and 2009, during the second and third phases of the OGLE project (OGLE-II and OGLE-III), the OGLE team found over 340,000 long-period variables (pulsating red giants) in the Magellanic System and Galactic bulge \citep{soszynski2009,soszynski2011,soszynski2013}. At least 100,000 of these objects should be potential LSP variables; however, for the vast majority of them the photometric amplitudes associated with the LSP changes are smaller than the amplitudes caused by the stellar pulsation. To avoid confusing the LSP modulation with other types of red giant variability (e.g. ellipsoidal variations or irregular dimming events due to dust obscuration), we retrieved from the OGLE catalogs only relatively large-amplitude, characteristic LSP light curves, similar to those shown in Figure~\ref{fig1}. Additionally, we carried out a limited search for LSP variables in the region of the sky not observed during the OGLE-III survey, but monitored during the fourth phase of the OGLE project (OGLE-IV) from 2010 to 2020.

\begin{figure*}
\centering
\includegraphics[width=17.7cm]{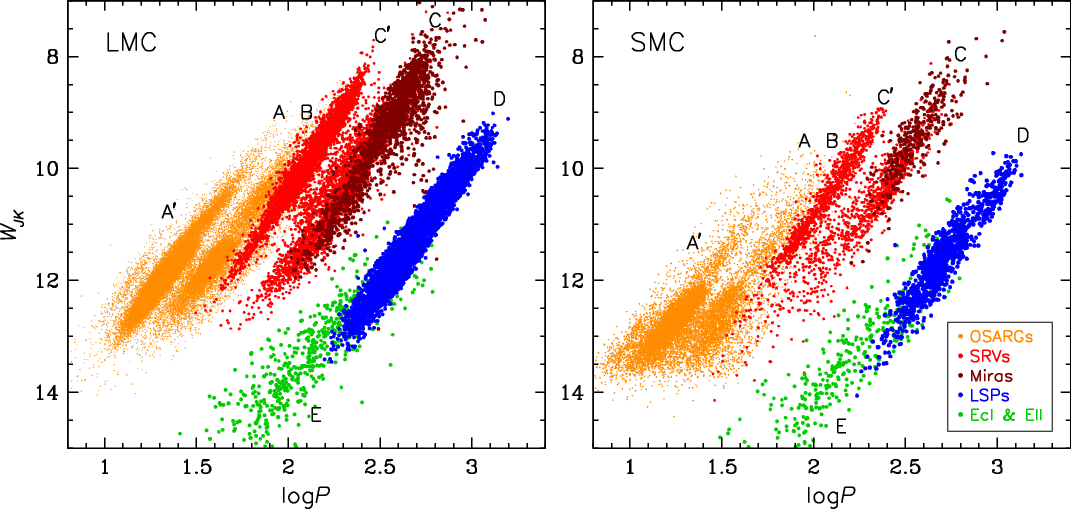}
\caption{Period--Wesenheit index diagrams for long-period variables in the LMC (left panel) and SMC (right panel). The NIR Wesenheit index is an extinction-free quantity defined as $W_{JK}=K_\mathrm{s}-0.686(J-K_\mathrm{s})$, where {\it J}- and $K_\mathrm{s}$-band magnitudes originate from the IRSF Catalog \citep{kato2007} or from the 2MASS Catalog \citep{cutri2003}. The period--luminosity sequences are labeled with letters according to the \citet{wood1999} nomenclature. Different colors of the points refer to different types of variable red giants: orange -- OGLE small-amplitude red giants (OSARGs; sequences A$'$, A, and B), red -- semiregular variables (SRVs; sequences C$'$ and C), brown -- Miras (sequence C), blue -- LSP variables (sequence D), and green -- eclipsing and ellipsoidal close binary systems containing red giants as one of the components (sequence E). Periods and classification of the stars were taken from the OGLE Collection of Variable Stars \citep{soszynski2009,soszynski2011}. \label{fig2}}
\end{figure*}

As a result, we found about 16,000 well-defined LSP variables: $\sim$9000 in the Milky Way and $\sim$7000 in the Magellanic Clouds. The OGLE-IV {\it I}-band light curves of these stars contain from a few hundred to more than 10,000 data points, depending on the field. Then, we fine-tuned the LSPs of our sample using the {\sc Tatry} code, which implements the multiharmonic analysis of variance algorithm \citep{schwarzenberg1996}. The classification of the LMC and SMC variables was additionally verified by the inspection of the near-infrared (NIR) period--luminosity diagrams for both galaxies, namely the $\log{P}$--$W_{JK}$ diagram (Figure~\ref{fig2}), where $W_{JK}=K_\mathrm{s}-0.686(J-K_\mathrm{s})$ is the extinction-free Wesenheit index, while {\it J}- and $K_\mathrm{s}$-band magnitudes were taken from the IRSF Magellanic Clouds Point Source Catalog \citep{kato2007} or (if not available) from the 2MASS All-Sky Catalog of Point Sources \citep{cutri2003}. Objects located outside sequence~D \citep{wood1999} in the period--luminosity diagram were removed from our list of LSP stars. The vast majority of the selected LSP variables are undoubtedly AGB stars, since no significant surplus of objects can be detected at luminosities below the RGB tip. In the next step, we extracted the mid-IR light curves of our sample.

\section{Mid-IR Photometry}

The Wide-field Infrared Survey Explorer \citep[WISE;][]{wright2010} is a 40~cm telescope onboard a satellite in a Sun-synchronous polar orbit around Earth. In 2010, the WISE telescope completed its primary mission by mapping the entire sky in four mid-IR bands -- W1, W2, W3, and W4 -- centered at 3.4, 4.6, 12, and 22~$\mu$m, respectively. After the depletion of the solid hydrogen cryostat, the spacecraft was placed into hibernation in early 2011 and reactivated in 2013. Since then, WISE has continued to survey the sky in the W1 and W2 channels. Every sky location is visited by the WISE telescope every six months and 12 or more independent exposures are made during each visit. The exception to this rule are regions around the ecliptic poles, which are observed much more frequently due to the polar trajectory of the WISE satellite. The South Ecliptic Pole is located in the outskirts of the LMC -- the area also covered by the OGLE survey, so a number of our LSP variables are located in this region.

We crossmatched our sample of LSP variables to the Near-Earth Object WISE Reactivation Mission \citep[NEOWISE-R;][]{mainzer2014} archive using a matching radius of $6''$ and we found counterparts for 89\% of objects. The NEOWISE-R light curves cover a time span of six years (2013--2019) and were obtained in two bands: W1 and W2. We decided not to include the photometry from the WISE primary mission (2010--2011), because we noticed that both datasets are not always compatible for individual stars, in particular there are magnitude shifts that would be difficult to account for.

The NEOWISE-R data points for a typical star are distributed over about 12 epochs at intervals of six months. Each epoch consists of 10--24 single-exposure observations obtained within one to a few days. We converted these individual measurements into fluxes, averaged (with outlier rejection) within each epoch, and converted back to the magnitude scale. As a result, we obtained W1 and W2 light curves containing typically 12 points (median value); however, about 2000 LSP variables, mostly in the LMC, have much-better-sampled light curves, consisting of several dozen or even over 100 points.

\begin{figure*}
\centering
\includegraphics[width=17.2cm]{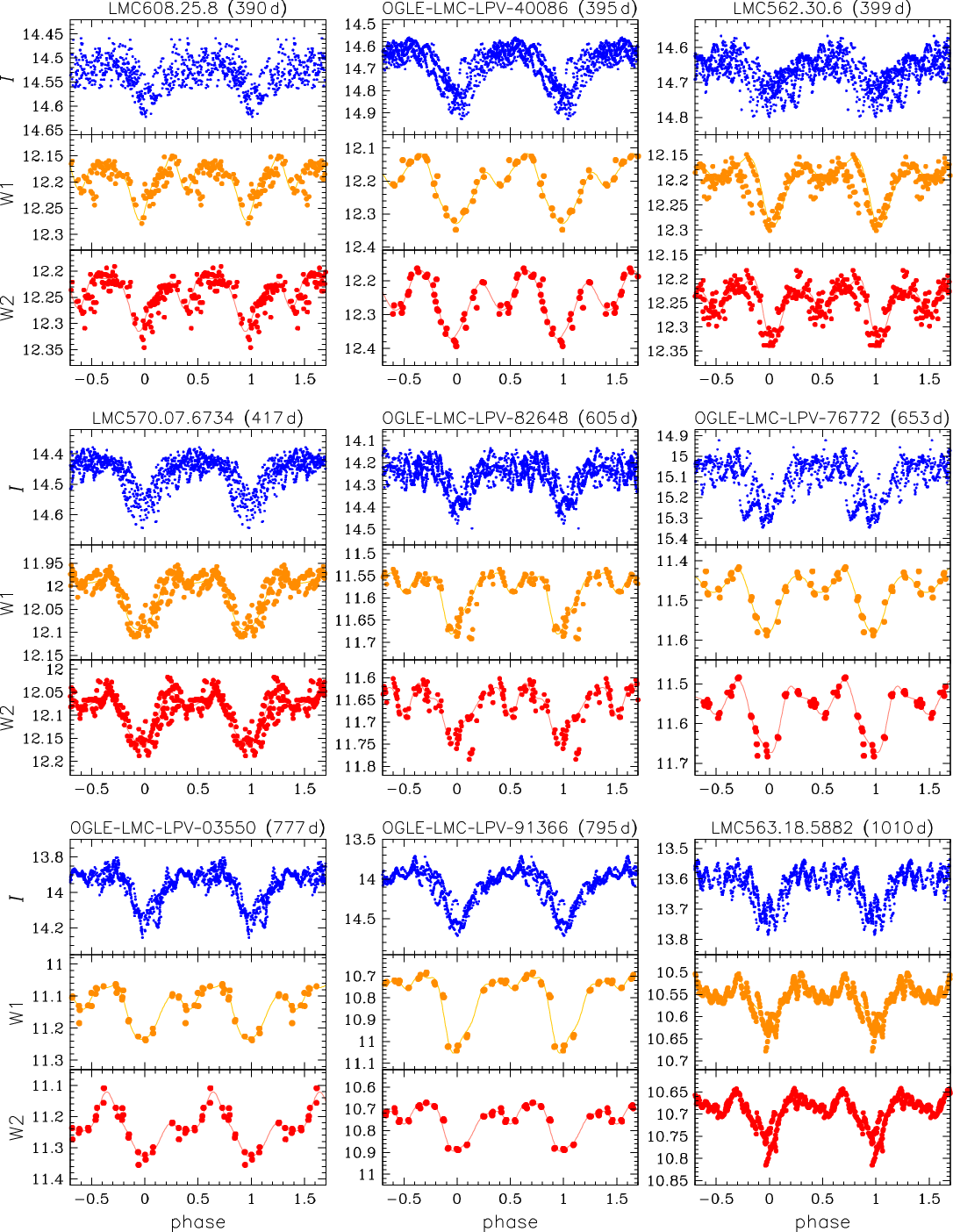}
\caption{Example light curves of LSP variables showing secondary minima in the mid-IR wavelengths. In each panel, blue, orange, and red points show OGLE-IV {\it I}-band, WISE W1, and WISE W2 photometry, respectively. Solid lines show the Fourier series fits to the mid-IR data. The LSPs are given in parentheses above the panels.\label{fig3}}
\end{figure*}

\section{Analysis of the Light Curves}

We inspected about 8500 phase-folded mid-IR light curves of the LSP stars with at least 12 epochs in each of the W1 and W2 bands. For a significant fraction of the sample, the phase coverage turned out to be too sparse to draw definite conclusions about the characteristics of the mid-IR time series. In particular, most stars with LSPs around 1, 1.5, or 2~yr had wide gaps in the folded light curves due to the six month intervals between the WISE observations. More than half of the examined LSP variables had peak-to-peak amplitudes of the mid-IR light curves below 0.1~mag, which in many cases was comparable to the noise caused by the (``short-period'') red giant pulsation and the errors of the WISE magnitudes. We also rejected some stars exhibiting irregular mean-magnitude variations and significant amplitude changes of the LSP modulation. As a result, we left with about 2500 objects for which the LSP variations could be easily recognized in the mid-IR data; however, a detailed analysis of the light-curve morphology was possible only for several hundred ($\sim$700) of them.

\begin{figure*}
\centering
\includegraphics[width=12.0cm]{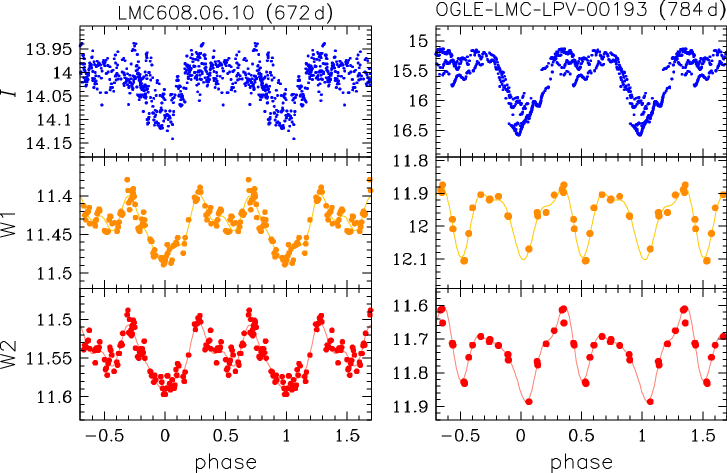}
\caption{Light curves of two LSP variables showing deep secondary minima in the mid-IR wavelengths (middle and lower panels) and shallow secondary minima in the {\it I}-band (upper panels).\label{fig4}}
\end{figure*}

The general characteristics of the WISE light curves turned out to be very similar to the characteristics of the optical time series, i.e. a broad triangular minimum lasting about half an LSP cycle is visible in the mid-IR data for the majority of large-amplitude LSP stars. The amplitudes of the W1 and W2 light curves are usually smaller than the amplitudes in the {\it I}-band; however, the $A_\mathrm{W1}/A_I$ and $A_\mathrm{W2}/A_I$ amplitude ratios range from about 0.3 to over 1 in various objects.

The primary minima in the optical and mid-IR wavelengths occur at the same time in the vast majority of LSP variables. We did not find any systematic phase lags between {\it I}-band and mid-IR light curves for the general sample of LSP stars, although positive or negative phase shifts can be detected in some individual cases. This is in contrast to the findings of \citet{takayama2020} who noticed systematic phase lags between the LSP light curves in the optical and NIR bands.

The most striking feature that distinguishes mid-IR from optical LSP light curves is the presence of secondary minima. In Figure~\ref{fig3}, we present {\it I-}, W1-, and W2-band light curves of nine LSP variables from the LMC -- all these stars exhibit well pronounced secondary minima in the W1 and W2 channels but this feature is not visible in the optical wavelengths. The LMC sources have the best-sampled WISE light curves (especially in the region of the sky around the South Ecliptic Pole), but we found such a behavior also in about 120 LSP variables in other environments: the bulge and disk of the Milky Way and the Small Magellanic Cloud.

As can be seen in Figure~\ref{fig3}, the secondary minima in the mid-IR light curves have different widths and depths in different LSP variables. Some stars show secondary minima as wide as the primary ones (half the LSP cycle), others exhibit secondary minima lasting only 0.2 of the LSP cycle or less. The secondary minima are generally much shallower than the primary ones, with the exception of very few LSP variables with secondary and primary minima of comparable depths. It is interesting that some of these extreme cases seem to exhibit shallow secondary minima also in the optical range. OGLE and WISE light curves of two such LSP variables in the LMC are shown in Figure~\ref{fig4}.

The efficiency of detecting secondary minima in the mid-IR light curves is strongly correlated with the phase coverage by the observations, as well as amplitudes, periods, and stability of the LSP light curves. Taking into account only relatively well-sampled W1 and W2 time series, we found about 350 candidates for LSP variables exhibiting the secondary minima. On the other hand, we also indicated a similar number of LSP variables with no distinct secondary minima in their mid-IR light curves. The remaining objects from our sample had too sparse photometry to draw strict conclusions about their IR light-curve morphology.

\section{Discussion}

As mentioned, the secondary minima in the mid-IR light curves are one of the predictions of the binary model of the LSPs in red giant stars. In this scenario, the red giant has a close-orbiting substellar or stellar companion surrounded by a comet-like dusty cloud. Hydrodynamical simulations of AGB binary systems \citep[e.g.][]{saladino2019,chen2020} show such spiral dusty structures formed by the matter lost by the giant due to stellar wind. The cloud absorbs the stellar radiation over a wide range of the electromagnetic spectrum and re-emits this energy at infrared wavelengths. When the red giant star is obscured by the cloud, we can observe the primary minima in all spectral ranges. When the dusty cloud is hidden behind the red giant, we can detect the secondary minima only in the mid-IR range. The shapes of the WISE light curves confirm that binarity is responsible for the LSP modulation observed in at least 30\% of pulsating red giants.

The amplitudes of the radial velocity curves measured by \citet{nicholls2009} for several dozen LSP variables suggest the presence of a brown dwarf or a very low-mass main-sequence star companion. Thus, the high prevalence of LSPs among long-period variables implies a large fraction of red giant stars with brown-dwarf companions. This is at odds with the so-called ``brown-dwarf desert'' -- a paucity of brown-dwarf companions relative to planets around main-sequence stars \citep{marcy2000,grether2006}. To explain this inconsistency, it should be assumed that the low-mass companions were initially planets that increased their mass accreting material from the stellar wind or via direct mass transfer. This means that the majority of LSP giants are orbited by objects formed from the planets and the total population of these variable stars can be used to study the populations of exoplanets in different regions of the Milky Way and in other galaxies.

The idea that planets orbiting red giants can be transformed into brown dwarfs or even low-mass stellar companions is not new \citep[e.g.][]{livio1984,retter2005}. It is also known that planets around red giant stars are on average more massive than those around main-sequence and subgiant stars \citep[e.g.][]{jones2014,niedzielski2015} which suggests that planets grow during the latest stages of their stellar hosts' evolution. According to \citet{wittenmyer2020}, about 7\% of main-sequence stars have a Jupiter-mass planet on wide orbits, but \citet{jones2021} estimated that $33^{+9}_{-7}$\% of giant stars host at least one Jovian planet within 5~au, which is in agreement with the observed occurrence rate of LSP stars among long-period variables. Note also that the majority of planetary nebulae do not have spherical symmetry, which is associated with the presence of stellar or substellar companions that survived the AGB phase of their stars \citep{decin2020}. Thus, the binary explanation of the LSP phenomenon is compatible with theoretical and observational arguments.

Some properties of the dusty cloud orbiting the red giant star can be learned from the shapes of the mid-IR light curves. The widths and depths of the secondary eclipses in relation to the primary eclipses depend on several factors: the structure and size of the obscuring cloud, its maximum temperature and temperature distribution inside the cloud, its transparency in various passbands, and the inclination of its orbit. As a result, the secondary eclipses may be too shallow to be detectable in some LSP stars, which we actually observe. Additionally, the pulsations of the host star, its irregular light variations, and sparsely sampled WISE light curves certainly reduce the effectiveness of our secondary eclipses' detection. However, the detailed calculations of the secondary eclipses' detection efficiency must be very complex and are beyond the scope of this paper.

Several of the LSP light curves shown in Figure~\ref{fig3} have secondary eclipses much narrower than the primary ones. Such a behavior can be interpreted as a result of specific density and temperature profiles of the orbiting dusty cloud. Simply, the densest and hottest part of the cloud that emits the bulk of mid-IR light may be significantly smaller than the whole cloud with its comet-like tail. The secondary eclipses can be observed only when this compact bright region of the cloud is hidden behind the star, which takes much less time than the primary eclipse, when the star is obscured by the entire cloud. In the case of the W2 light curve of OGLE-LMC-LPV-03550 (lower left panel of Figure~\ref{fig3}), the secondary eclipse seems to begin before the primary eclipse is ended. This may happen when the dusty tail is exceptionally long.

The secondary minima are most often visible in both W1 and W2 light curves, and the W2 eclipses are usually deeper. Relative depths of the secondary eclipses in both mid-IR channels can give us a hint about the temperature of the dusty cloud. For this purpose, we performed a simple exercise. Using the PHOEBE2 package \citep{prsa2005}, we derived a series of synthetic light curves of eclipsing binary systems containing a $T=3500$~K primary (red giant) and a secondary with temperatures ranging from 1000 to 3500~K (equivalent of the dusty cloud). ATLAS9 \citep{castelli2003} stellar atmospheres were used for the primary. Since the effective temperature values of the secondary fall outside the ATLAS9 range, we used the blackbody stellar atmospheres for this component. Obviously, hotter dusty clouds produce deeper secondary eclipses in both WISE filters, but we also found that the temperature of the cloud is strongly correlated with the W2-to-W1 ratios of the eclipse depths. Namely, for the lowest checked temperature (1000~K), the W2 amplitudes are larger by a factor of 1.5 than the W1 amplitudes, while for the temperatures around 3500~K, the depths of the secondary minima in the W1 and W2 bands are nearly equal. The W2-to-W1 ratios of the secondary eclipse depths derived for the LSP light curves shown in Figure~\ref{fig3} revealed that the temperature of the dusty cloud in most of these objects is in the range 1000--1500~K.

In most LSP light curves, the secondary minima are located symmetrically between the primary minima (at the phase around 0.5), which indicates the circular orbits of the low-mass companions. The tidal circularization theory \citep{zahn1977,soker2000} predicts that the majority of close binary systems containing red giant components should have low eccentricities, because the typical circularization timescales are only several thousand years \citep{nicholls2010}. However, we also found examples of the LSP variables exhibiting clearly nonsymmetrically positioned secondary minima, which suggests eccentric orbits. We estimate that such asymmetric mid-IR light curves appear in about 10\%--15\% of the LSP giants with detectable secondary minima. It is striking that a similar fraction of eccentric systems was found by \citet{soszynski2004} in the sample of ellipsoidal red giants -- the close binary system populating sequence~E in the PL diagram.

\section{Conclusions}

\citet{wood2004} stated that LSPs ``represent the only form of unexplained, large-amplitude stellar variability known at this time.'' This is no longer true. We demonstrated that a large fraction of LSP variables exhibit secondary minima in their mid-IR light curves, which proves that the physical mechanism responsible for the LSP modulation is the obscuration by a dusty cloud orbiting the red giant with a substellar or stellar companion. In this scenario, this object is a former planet that accreted a significant amount of mass from the envelope of its host star and appeared as a brown dwarf or a very low-mass main-sequence star. Thus, the phenomenon known for many decades \citep{oconnell1933,payne1954,houk1963} seems to be a tracer of extrasolar planets.

Careful examination of the mid-IR and optical light curves can unveil properties of the LSP variables. We showed that the majority of LSP systems have circular orbits, as is expected from the tidal circularization theory. Note that the observed nonsinusoidal shapes of the radial velocity curves \citep{hinkle2009,nicholls2009} are not in conflict with this finding \citep{soszynski2014}. Based on the secondary eclipse widths and depths in the two mid-IR passbands, we demonstrated that the orbiting dusty cloud has a nonhomogeneous structure and its temperature may reach $\sim$1500~K in the densest part. For lower orbital inclinations or cooler dusty clouds the secondary eclipses may not be visible in the WISE photometry, which also happens in some LSP stars.

\acknowledgments

We thank the anonymous referee for the constructive comments that helped us to improve the manuscript. We are grateful to Andrzej Niedzielski for helpful suggestions that improved the paper. This work has been supported by the National Science Centre, Poland, grant MAESTRO No. 2016/22/A/ST9/00009. The OGLE project has received funding from the Polish National Science Centre grant MAESTRO No. 2014/14/A/ST9/00121. This publication makes use of data products from the Near-Earth Object Wide-field Infrared Survey Explorer (NEOWISE), which is a joint project of the Jet Propulsion Laboratory/California Institute of Technology and the University of Arizona. NEOWISE is funded by the National Aeronautics and Space Administration.

\end{document}